\documentclass{jpp}
\usepackage{graphicx}

\usepackage[utf8]{inputenc}
\usepackage[T1]{fontenc}
\usepackage{amsmath}
\usepackage{graphicx}
\usepackage{dcolumn}
\usepackage{bm}
\usepackage{booktabs}
\usepackage{siunitx}
\usepackage[utf8]{inputenc}
\usepackage[T1]{fontenc}
\usepackage{mathptmx}
\usepackage{etoolbox}
\usepackage{hyperref}
\hypersetup{
    colorlinks = true,
    linkcolor = blue,
    citecolor = blue,
}

\shorttitle{3D collisional magnetized bow shocks in pulsed-power-driven plasma flows}
\shortauthor{R. Datta, D. R. Russell, I. Tang, et al.}

\title{The structure of 3D collisional magnetized bow shocks in pulsed-power-driven plasma flows}

\author{R. Datta\aff{1},
  D. R. Russell\aff{2},
  I. Tang\aff{2},
  T. Clayson\aff{2},
  L. G. Suttle\aff{2},
  J. P. Chittenden\aff{2},
  S. V. Lebedev\aff{2},
 \and  J. D. Hare\aff{1}
 \corresp{\email{jdhare@mit.edu}}}

\affiliation{\aff{1}Plasma Science and Fusion Center, Massachusetts Institute of Technology, Cambridge MA02139, USA
\aff{2}Blackett Laboratory, Imperial College London, London SW7 2BW, UK}

\begin{document}

\maketitle

\newcommand{\ra}[1]{\renewcommand{\arraystretch}{#1}}

\begin{abstract}
We investigate 3D bow shocks in a highly collisional magnetized aluminum plasma, generated during the ablation phase of an exploding wire array on the MAGPIE facility (1.4 MA, 240 ns). Ablation of plasma from the wire array generates radially diverging, supersonic ($M_S \sim 7$), super-Alfvénic ($M_A > 1$) magnetized flows with frozen-in magnetic flux ($R_M \gg 1$). These flows collide with an inductive probe placed in the flow, which serves both as the obstacle that generates the magnetized bow shock, and as a diagnostic of the advected magnetic field. Laser interferometry along two orthogonal lines of sight is used to measure the line-integrated electron density. A detached bow shock forms ahead of the probe, with a larger opening angle in the plane parallel to the magnetic field than in the plane normal to it. Since the resistive diffusion length of the plasma is comparable to the probe size, the magnetic field decouples from the ion fluid at the shock front and generates a hydrodynamic shock, whose structure is determined by the sonic Mach number, rather than the magnetosonic Mach number of the flow. 3D simulations performed using the resistive magnetohydrodynamic (MHD) code GORGON confirm this picture, but under-predict the anisotropy 
observed in the shape of the experimental bow shock, suggesting that non-MHD mechanisms may be important for modifying the shock structure.

\end{abstract}

\section{\label{sec:intro} Introduction}

 Shocks are a fundamental process of kinetic energy dissipation in space and astrophysical plasmas. Many astrophysical objects generate high Mach number flows which interact with ambient media or planetary obstacles to generate strongly radiating plasma shocks. Some examples include extragalactic and relativistic jets from radio galaxies \citep{Miley1980, Smith1981,  Duncan1994, Choi2007}, Herbig-Haro jets from young stellar objects (YSOs) \citep{Hartigan1990, Smith2003, Smith2012}, and shocks in core-collapse supernovae and supernova remnants \citep{Chevalier1982,Kifonidis2003}. Many astrophysical environments also contain dynamically significant magnetic fields, that result in the formation of magnetized plasma shocks which exhibit physics not found in unmagnetized hydrodynamic shocks \citep{DeSterck1998,Dursi2008}.
 
 Laboratory astrophysics experiments at high-energy-density-plasma (HEDP) facilities have provided key insight into the physics of plasma shocks \citep{Remington2006,Lebedev2019}. Laser plasma experiments have been used extensively to study physics relevant to astrophysical shocks, such as the evolution of hydrodynamic shock instabilities \citep{Remington1997,Kane1999}, and the interaction of shocks and jets with low-density ambient media \citep{Drake1998, Robey2002, Foster2005}. More recent experiments have investigated the formation of magnetized shocks generated from the interaction of laser-driven plasma pistons in the presence of externally applied magnetic fields
 \citep{Schaeffer2015,Schaeffer2017,Liao2018,Schaeffer2022,Levesque2022}.
 
 Laser-driven magnetized shock experiments have typically focused on the study of collisionless shocks --- shocks in which dispersive two-fluid effects, as opposed to collisional dissipation, facilitate the transition at the shock front. \citep{Goldstein1984,Schaeffer2015,Schaeffer2017,Schaeffer2022}. In contrast, pulsed-power-driven plasmas can be used to generate magnetized shocks in the highly-collisional regime. Pulsed-power machines generate plasma by applying a large current ($\sim 1-30$ MA) to a load, typically an array of thin wires, over a short time ($\sim 100-\SI{300}{\nano \second})$. The ablation of plasma from wire arrays generates highly collisional ($\lambda_{ii}/L \ll 1$), supersonic and super-Alfvénic upstream flows with frozen-in magnetic flux ($R_M \equiv VL/\bar{\eta} \gg 1$) \citep{Burdiak2017,Hare2018,Suttle_2019}. Here, $\lambda_{ii}$ is the ion-ion mean free path, $L$ is the characteristic size of the plasma, $V$ is the characteristic bulk flow velocity, and $\bar{\eta}$ is the magnetic diffusivity. Pulsed-power has been used extensively to study physics relevant to supersonic astrophysical jets, such as the interaction of plasma jets with neutral gases \citep{Suzuki-Vidal2012, Suzuki-Vidal2013}, the fragmentation of radiatively-cooled bow shocks in counter-propagating jets \citep{Suzuki-Vidal2015}, and the structure of magnetized oblique shocks \citep{Swadling2013}, planar shocks \citep{Lebedev2014}, and quasi-2D bow shocks \citep{Ampleford2010, Bott-Suzuki2015, Burdiak2017}.
 
Previous experimental work on collisional shocks has shown that magnetic fields can modify the structure of collisional plasma shocks in two primary ways --- either via the generation of magnetohydrodynamic (MHD) shocks, \citep{Lebedev2014,Russell2022} or via magnetic flux pile-up \citep{Burdiak2017,Suttle_2019}. In MHD shocks, the magnetic field can change discontinuously across the shock front, resulting in complicated anisotropic shock structures, whose morphology depends not only on the velocity of the upstream flow relative to the three MHD wave speeds (fast, slow, and Alfvén waves), but also on the orientation of the upstream magnetic field relative to the shock front \citep{DeSterck1998,DeSterck2000,Verigin2001,goedbloed_keppens_poedts_2010}. Recent experiments have also investigated the role of resistive dissipation in MHD shock transitions through the formation of sub-critical collisional shocks \citep{Russell2022}.

A second effect that can modify the structure of shocks in magnetized plasmas is magnetic flux pile-up \citep{Dursi2008}. Magnetic field lines advected by the plasma accumulate and drape around conducting obstacles. The draping of field lines around cylindrical obstacles has been shown to modify the structure of quasi-2D bow shocks, resulting in wider shocks with a larger stand-off distance when the obstacle axis is oriented perpendicular to the field, than when it is parallel to the field \citep{Burdiak2017}. Although quasi-1D and quasi-2D shocks in magnetized plasmas have been examined extensively, full 3D shocks in magnetized pulsed-power plasmas have received less attention.

In this paper, we show that bow shocks around small 3D obstacles in a supersonic ($M_S = V_{\text{flow}}/C_S \sim 7$) and super-fast magnetosonic ($M_{FMS} = V_{\text{flow}}/V_{FMS} > 1$) plasma exhibit a fully 3D structure, with a larger shock opening angle in the plane parallel to magnetic field, than in the plane normal to the field. At large length scales, comparable to the size of the plasma, magnetic flux is frozen into the flow ($R_M \gg 1$); however, at length scales comparable to the resistive diffusion length $l_\eta$ ($R_M \sim 1$), the magnetic field decouples from the plasma. For obstacles of size comparable to the resistive diffusion length, the breaking of frozen-in flux due to resistive diffusion results in hydrodynamic bow shocks, where the magnetic field remains continuous across the shock front.
The obstacle used in these experiments is an inductive probe, which not only generates the shock, but also measures the post-shock magnetic field. Although inductive probes are widely used in HEDP experiments, \citep{Everson2009,Pilgram2021,Suttle_2019} their perturbative nature leads to questions about how reliably they can reconstruct the magnetic field in plasma flows. These experiments additionally tackle this question by careful comparison between numerical simulations and experimental data.

\section{\label{sec:setup}Experimental and Diagnostic Setup}

\begin{figure}
\includegraphics[page=1,width=1.0\textwidth]{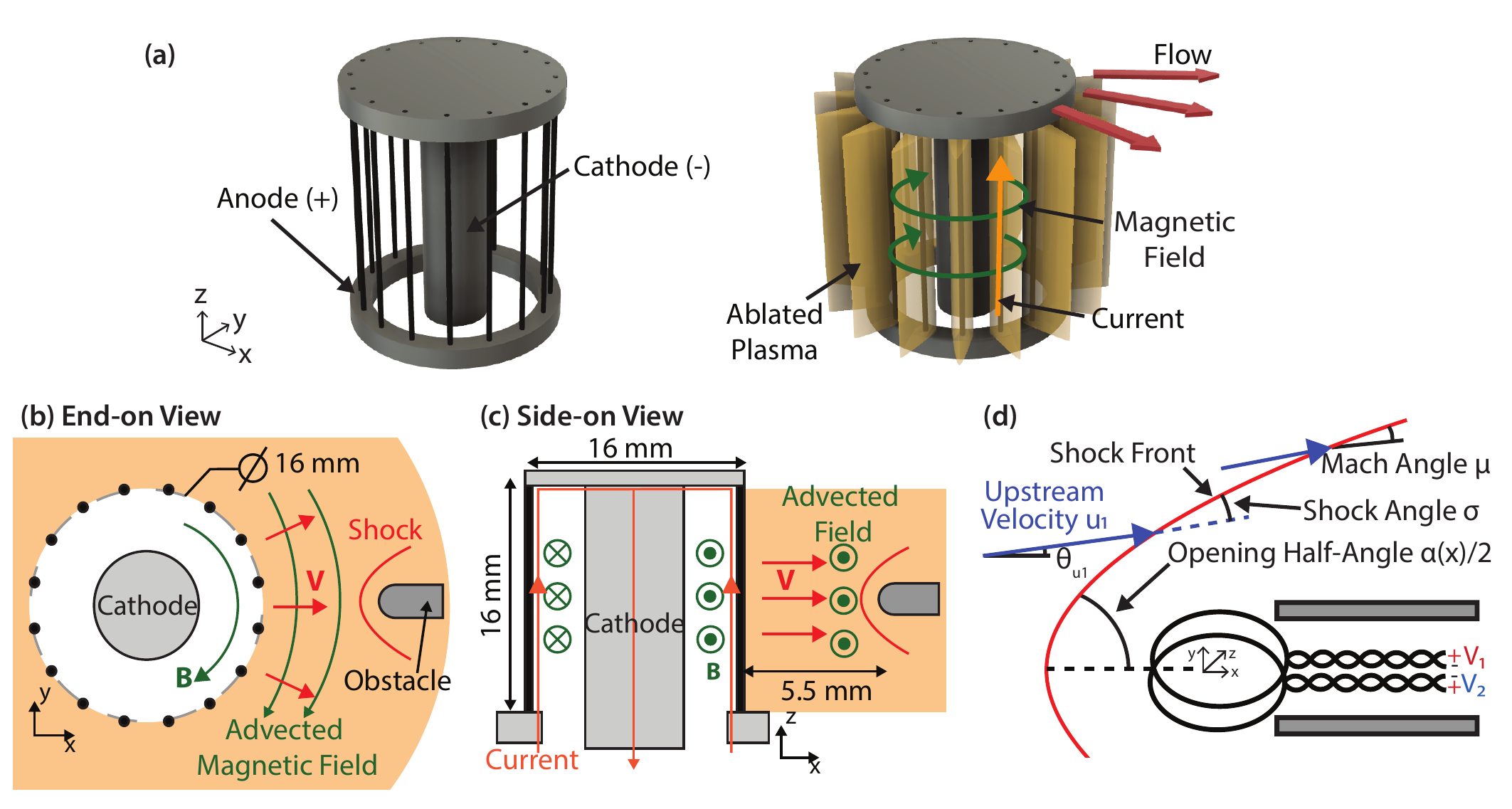}
\centering
\caption{(a) 3D representation of an exploding wire array (b) End-on (xy-plane) view of the experimental geometry, showing a cylindrical array of 16 equally-spaced \SI{30}{\micro\metre} Al wires around a central cathode. An inductive probe serves as the obstacle and is placed $\sim 5.5$ mm from the array surface. 
(c) Side-on (xz-plane) view of the experimental geometry. (d) Schematic showing bow shock geometry observed ahead of the inductive probe in the end-on view. The shock front is represented by the red solid line.}
\label{fig:exp_setup_bow}
\end{figure}

\subsection{Load Hardware}
Figure \ref{fig:exp_setup_bow} illustrates the load and the experimental setup. The load consists of a cylindrical array of 16 equally-spaced, \SI{30}{\micro\metre} diameter aluminum wires (California Wire Company) positioned around a central 5 mm diameter stainless steel cathode. The array diameter and the array height are both \SI{16}{\milli\metre}. The current pulse (1.4 MA peak current, 240 ns rise time) is generated using the MAGPIE machine at Imperial College London \citep{Mitchell1996}.

When current flows through the wires, the wires heat up resistively, and the wire material vaporizes and ionizes to create low-density coronal plasma surrounding the dense wire cores. The current, which travels along the wires, is mostly concentrated within a thin skin region containing the coronal plasma immediately around the stationary wire cores. The global magnetic field points azimuthally inside the array, and rapidly drops to zero outside the array \citep{Velikovich2002}. The global ${\bf j \times B}$ force, therefore, accelerates the coronal plasma radially outwards, and the ablated plasma streams supersonically into the vacuum region outside the array. 


The magnetic Reynolds number, calculated with the characteristic experimental length scale $L \sim \SI{1}{\centi \metre}$, is large  $R_M \sim 10-100$ \citep{Suttle_2019}, so magnetic flux is frozen into the flow, and the ablating plasma advects part of the global field with it. The ion-ion mean free path of the ablating plasma is also small ($\lambda_{ii} \sim 10^{-3} \SI{}{\milli \metre}$), so the plasma is highly-collisional \citep{Suttle_2019}. The flow velocity in similar setups is typically supersonic ($M_S \sim 3-5$), super-Alfvénic ($M_A \sim 2$), and super-fast magnetosonic ($M_{FMS} \sim 2$) \citep{Burdiak2017,Russell2022}. The collision of these supersonic outflows  with the obstacle generates a detached bow shock. 
The adiabatic index of the plasma (ratio of specific heats) is an important quantity that affects the shock physics. In HED plasmas, contributions to internal energy and pressure due to Coulomb interactions, ionization, and excitation processes can make the effective adiabatic index lower than that of an ideal gas ($\gamma_{\text{ideal}} = 5/3$). In HED plasmas with characteristic electron density $n_e \sim 1 \times 10^{18} \SI{}{\per \centi \metre \cubed}$ and electron temperature $T_e \sim 10$ eV, the typical value of  $\gamma_\text{{eff}} \sim 1.1-1.2$ \citep{Drake2006,Swadling2013}. 

\subsection{Diagnostic Setup}

In contrast to previous experimental work \citep{Lebedev2014, Burdiak2017, Suttle_2019, Russell2022}, an inductive (`b-dot') probe serves as the obstacle, and is positioned $5.6 \pm \SI{0.3}{\milli \metre}$ from the wires. In addition to generating the bow shock, the probe also measures the advected magnetic field. The probe consists of two $\sim \SI{0.5}{\milli \metre}$ diameter loops of oppositely-wound single-turn enamel-coated copper wire, threaded through a $\sim 1$ mm diameter thin-walled steel tube. The voltage response of the probe can have two contributions --- one due to the time-varying magnetic flux through the loop, and another electrostatic component due to the coupling of stray voltages from the pulsed-power generator. Having two oppositely-wound loops provides a differential measurement, allowing us to combine the raw, unintegrated signals from both loops, and isolate the contribution of the time-varying magnetic flux \citep{Suttle_2019, Datta2022}. We position the inductive probe to measure the azimuthal magnetic field --- the normals to the surfaces of the loops lie along the magnetic field. The voltage signal from the probe is proportional to the rate of change of the magnetic field $V = \dot{B}A_\text{eff}$. To determine the magnetic field strength at the probe, we numerically integrate the voltage signal in time. The inductive probe was calibrated before use in the experiment to determine its effective area  $A_{\text{eff}} = 0.30 \pm \SI{0.01}{\milli\metre\squared}$. This was done by placing the probe within the known magnetic field generated by a Helmholtz coil driven by $\sim$ 1 kA time-varying current. The current was measured using a calibrated Pearson coil, and a Biot-Savart solver was used to calculate the magnetic field from the known coil geometry and the measured current.

A second inductive probe ($A_{\text{eff}} = 0.28 \pm \SI{0.01}{\milli\metre\squared}$), identical to the one described above, was also positioned at the same radial distance from the array, but at a different azimuthal location. This probe lies outside the field-of-view of our imaging diagnostic, but provides a second measurement of the magnetic field at the same radial position. In addition to the two probes placed in the flow, a single-loop b-dot probe is positioned in a recess near the current feed for the load. This probe is uncalibrated, and monitors the the current delivered to the load.

We use a Mach-Zehnder interferometer to visualize the plasma flow and the bow shock around the first inductive probe. The interferometry setup simultaneously provides both end-on (Figure \ref{fig:exp_setup_bow}b) and side-on views (Figure \ref{fig:exp_setup_bow}c) of the experimental setup. The end-on interferometer, which provides an axially integrated view of the experimental setup in the x-y plane, is illuminated using a $\SI{532}{\nano \metre}$ pulsed Nd-YAG laser (EKSPLA SL321P, \SI{500}{\pico \second}, \SI{100}{\milli \joule}). The side-on interferometer provides a line-integrated (along the y-direction) view of the x-z plane, and is illuminated using a 1053 nm Nd:Glass laser (\SI{1}{\nano \second}, \SI{1}{ \joule}). Both laser beams are expanded to provide a $\sim \SI{20}{\milli \meter}$ field-of-view. We combine the probe and reference beams at the CCD of a Canon EOS 500D DSLR camera. When the probe beam propagates through the plasma, the resulting phase accumulated by the beam distorts the fringe pattern, and introduces a spatially-varying fringe shift \citep{Hutchinson2002,Swadling2013}, which we use to reconstruct the phase difference between the probe and reference beams, and to determine the spatially-resolved line-integrated electron density \citep{Hare2019}. 

\section{Results}

\begin{figure}
\includegraphics[width=1.0\textwidth,page=2]{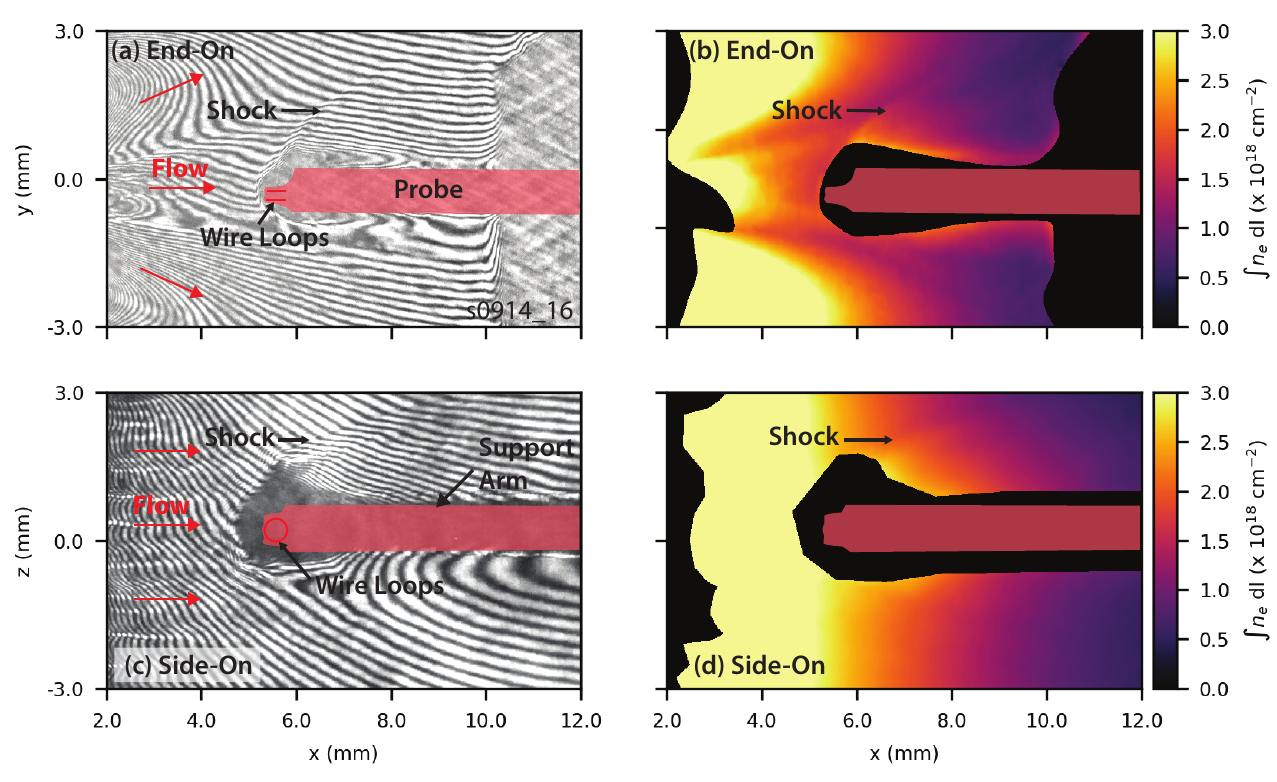}
\centering
\caption{(a) End-on raw interferogram at 300 ns after current start using a Mach-Zehnder interferometer with a 532 nm laser. The red shaded region represents the silhouette of the obstacle from the background interferogram recorded before start of the experiment.
(b) End-on line-integrated electron density map determined from interferometry.
(c) Side-on raw interferogram at 300 ns after current start using a 1053 nm laser.  
(d) Side-on line-integrated electron density map determined from interferometry. Regions in black near the obstacle and the wire array surface represent locations where the probing beam is lost. 
}
\label{fig:interferometry}
\end{figure}

\subsection{Bow Shock Morphology}
\label{sec:results}

Figure \ref{fig:interferometry} shows the end-on (x-y) and side-on (x-z) raw interferograms, together with the line-integrated electron density maps at $t = \SI{300}{\nano \second}$ after current start. We use a coordinate system centered at the intersection of the obstacle axis and the array surface throughout this paper. Note that the magnetic field lines lie tangential to the end-on plane, and normal to the side-on plane. A bow shock, characterized by a curved discontinuity in electron density, is visible in both end-on and side-on images. The bow shock appears more distinct on the top of the probe in both views, because the shock front is almost parallel to the fringes under the probe, so the fringes appear relatively undisturbed. On the other side of the probe, where the fringes are at an angle to the shock, the shock front appears more prominent. Nevertheless, we expect the shock front to be axisymmetric about the obstacle axis, due to the symmetry of the upstream flow.

  Figures \ref{fig:interferometry}b $\&$ \ref{fig:interferometry}d show that the electron density is high near the surface of the wire array and decreases with distance from the array, as expected due to the radially-diverging nature of the outflows. In the end-on plane, the upstream flow exhibits significant modulation in the azimuthal direction. This modulation results from the supersonic collision of adjacent azimuthally-expanding jets ablating from the wire cores, which forms standing oblique shocks periodically distributed between the wires \citep{Swadling2013}. Due to the oblique shocks, we expect the Mach number of the upstream flow in the end-on plane to also exhibit periodic modulation. In comparison, the upstream flow exhibits little modulation in the z-direction (Figure \ref{fig:interferometry}d). 

We define the shock opening half-angle $\alpha /2$ to be the angle between the obstacle axis and the shock front (see Figure \ref{fig:exp_setup_bow}d). Figure \ref{fig:interferometry} clearly shows the anisotropy in the shock structure --- the shock is wider in the end-on plane (i.e. it has a larger opening angle) than in the side-on plane. The shock angle $\sigma$ is the angle between the upstream velocity vector and the shock front (see Figure \ref{fig:exp_setup_bow}d). If the upstream velocity vector ${\bf u_1}$ makes an angle $\theta_{u1}$ with respect to the horizontal, the shock angle then becomes $\sigma = \alpha/2 - \theta_{u1}$.

To determine the opening-half angle from the observed bow shock geometry, we trace the shock front and fit a curve $s(x_s,y_s)$ to it. The opening half-angle is then simply $\alpha(x_s)/2 = \tan^{-1}(dy_s/dx_s)$. To calculate the shock angle $\sigma$, we must account for the direction of the upstream velocity. In the side-on plane, the projection of the upstream velocity only has a component along the x-direction, i.e. $\theta_{u1} = 0$ (see Figure \ref{fig:interferometry}), so the opening half-angle and the shock angle are equal in this plane. In the end-on view, however, the velocity vector makes a non-zero angle with the horizontal due to the radially diverging nature of the flow. We assume that the upstream velocity propagates radially outwards with respect to the array center i.e. ${\bf u_1} = u_1 {\bf \hat{e}_r}$. The velocity vector then makes an angle $\theta_{u1} = \tan^{-1}(y_s/x_s)$ to the horizontal. 

In bow shocks, the shock angle varies continuously from \SI{90}{\degree} at the nose of the obstacle to the Mach angle $\mu$ asymptotically far away from the obstacle, where the bow shock constitutes an infinitesimally weak Mach wave \citep{anderson_2001}. From our interferometry images (Figure \ref{fig:interferometry}), we observe that the shock opening half-angle asymptotically approaches $\alpha/2 \rightarrow $\SI{30}{\degree} and $\alpha/2 \rightarrow $ \SI{7}{\degree} far away from the obstacle in the end-on and side-on views respectively. Accounting for the direction of the upstream velocity, as described in the previous paragraph, the Mach angles are $\mu \approx 11 \pm$ \SI{0.5}{\degree} (end-on) and $\mu \approx 7 \pm$ \SI{0.5}{\degree} (side-on). The Mach angle is $\sim \SI{4}{\degree}$ higher in the end-on plane. We discuss this difference between the end-on and side-on Mach angles in \S \ref{sec:discussion}. 

\subsection{Magnetic Field Measurements}

\begin{figure}
\includegraphics[width=1.0\textwidth,page=3]{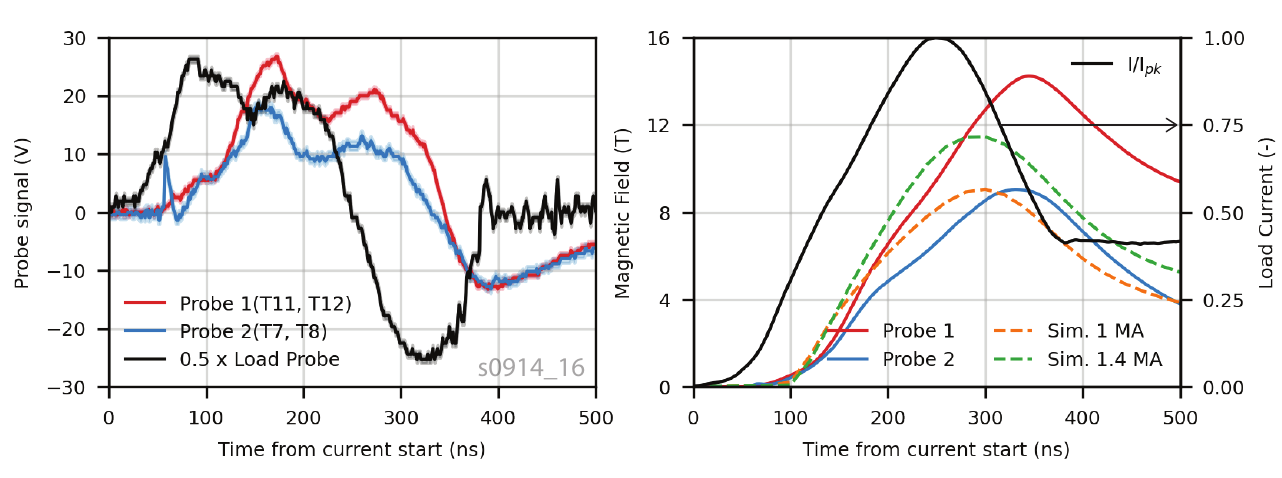}
\centering
\caption{(a) Signal from load probe and inductive probes placed in the flow. The probe signals are displaced in time relative to the signal from the load probe, showing that the magnetic field is advected by the flow. (b) Time-resolved load current and magnetic field. Peak field value is $\sim 9-14$ T and occurs at $\sim 340$ ns after current start. }
\label{fig:bdot_measuremnts}
\end{figure}

Figure \ref{fig:bdot_measuremnts}a shows the voltage signals from the two inductive probes placed in the flow, as well as from the probe monitoring the current in the load. The signal from the load probe is proportional to the time rate of change of the current in the wire array, and exhibits a characteristic `double-bumped' structure with a larger peak at $\sim 85$ ns followed by a smaller peak at $\sim 180$ ns. These peaks are caused by voltage reflections from impedance mismatches within the transmission lines of the pulsed-power machine \citep{Mitchell1996}. The rise time of the load current is $\sim \SI{250}{\nano \second}$. Due to a lack of calibration information, we only show the shape of the current waveform rather than its magnitude. Rogowski coil measurements around return posts in experiments with similar loads show that MAGPIE consistently delivers a 1.4 MA peak current \citep{Lebedev2014, Burdiak2017, Hare2018}.
The probes in the flow reproduce the shape and characteristic features of the signal at the load, showing that the magnetic field is frozen-into the flow, and that the magnetic field from the inside of the array is advected to the outside by the ablating plasma \citep{Burdiak2017}. 

Figure \ref{fig:bdot_measuremnts}b shows the load current and the advected post-shock magnetic field. The current signal has been normalized using its peak value. The load current and the advected magnetic field are determined by numerically integrating the load probe signal and the flow probe signals respectively. The load current and the advected magnetic field have similar shapes, again confirming frozen-in flux. The two probes, although placed at the same radial location and calibrated before use,  measure peak magnetic field strengths of $14$ T and $9$ T respectively. A possible source of this discrepancy is misalignment of the probe normal with respect to the magnetic field vector. However, a large misalignment ($\sim \SI{50}{\degree}$) would be necessary to account for the observed difference. Since the rotational position of the probes was adjusted to return the maximum signal during calibration, a large misalignment of this scale is unlikely.  

\section{3D resistive MHD Simulations in GORGON}

\begin{figure}
\includegraphics[width=1\textwidth,page=4]{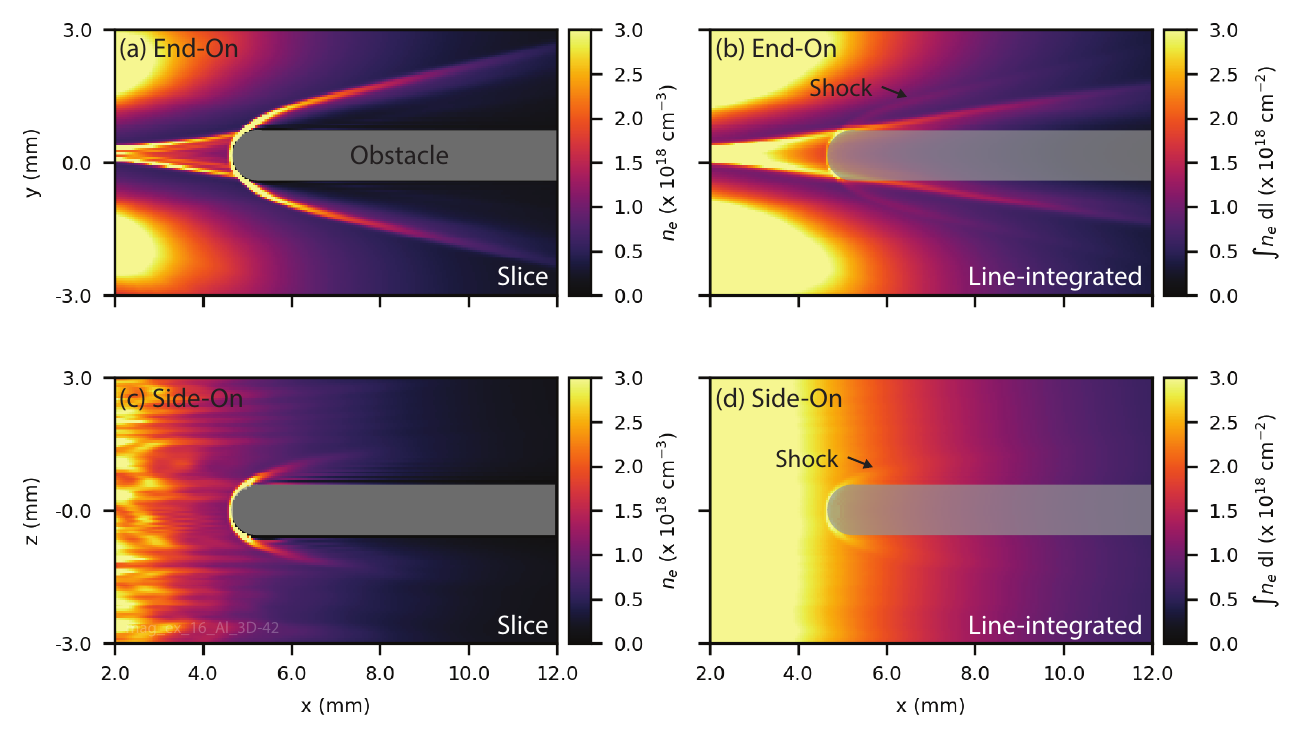}
\centering
\caption{
(a) End-on slice of electron density at 300 ns from 3D resistive MHD GORGON simulation of the experimental geometry.
(b) End-on line-integrated electron density at 300 ns from 3D resistive MHD GORGON simulation of the experimental geometry.
(c) Side-on slice of simulated electron density at 300 ns.
(d) Side-on line-integrated electron density at 300 ns from simulation. In each image, the plasma flow is from the left to the right.
}
\label{fig:sim_endon}
\end{figure}

To better understand these experiments, we use GORGON,  a 3D (cartesian, cylindrical, or polar coordinate) Eulerian resistive MHD code with van Leer advection, and separate energy equations for ions and electrons \citep{Chittenden2004, Ciardi2007}. Many radiation loss and ionization models can be implemented in GORGON. Here, we use a simple volumetric recombination-bremssthralung model \citep{Huba2013} modified with a constant multiplier to account for line radiation, and an LTE Thomas-Fermi equation-of-state (EOS) to determine the ionization level. For the simulated plasma and bow shock, the cooling time due to radiative recombination and bremssthralung losses $\tau_{\text{cool}} \sim p / P_{\text{rad}}$ is small compared to the hydrodynamic time scale $\tau_H \sim L/C_S$ of the plasma, and the simulation results remain largely unchanged when the radiation multiplier is set to unity. Here, $p$ is the thermal pressure, $P_{\text{rad}}$ is the radiative power loss, and $C_S$ is the sound speed. We simulate an exploding wire array with the same geometric dimensions, wire material, and wire thickness, as in the experimental setup. The current pulse applied to the load was determined from a sum-of-sines fit to the integrated signal of a Rogowski coil around a return post from a different MAGPIE shot with a similar load. We use a peak current of $I_{\text{pk}}  = 1$ MA, instead of $I_{\text{pk}} = 1.4$ MA, because the simulated density better matches the experimentally observed electron density for the former case. For $I_{\text{pk}} = 1.4$ MA, both the simulation and the rocket model \citep{Lebedev2001} predict that the wires explode before 300 ns after current start. Since the wires appear to still be in the ablation stage at 300 ns from the experimental images, we believe that the current delivered to the load is lower than what is measured by the Rogowski. The simulation domain is a cuboid with dimensions $51.2 \times 50.4 \times 38$ \SI{}{\milli\metre\cubed}. The initial mass in the wires is distributed over $3 \times 3$ grid cells pre-expanded wire cores. We place a resistive ($\eta = 0.1 \Omega \SI{}{\metre}$) cylindrical obstacle of diameter 1 mm at 5.5 mm from the array edge. The cylindrical obstacle mimics the inductive b-dot probe in the experiment, and is aligned parallel to the x-axis. The leading edge of the cylinder is a non-conducting sphere of diameter 1 mm. The simulations are performed with a grid size of $ \SI{50} {\micro \metre}$. Our convergence study shows that this resolution is adequate to achieve convergence.



Figures \ref{fig:sim_endon}a and \ref{fig:sim_endon}c show the end-on and side-on slices of the simulated electron density through the obstacle mid-plane at $t = \SI{300}{\nano \second}$ after current start, and Figures \ref{fig:sim_endon}b and \ref{fig:sim_endon}d show the end-on and side-on line-integrated electron density at the same time. A detached bow shock is visible ahead of the obstacle in the electron density slices and the line-integrated maps. In the line-integrated electron density maps, the shock front appears `muted', similar to what we observe in the experimental image, because line-integrating obfuscates the density jump at the 3D shock front. 
\section{Discussion of Results}
\label{sec:discussion}

\subsection{Bow Shock Structure}

The structure of shocks is closely related to the propagation velocity of linear perturbations in a given medium \citep{goedbloed_keppens_poedts_2010,Kundu-2012}. For isotropic waves, we can use the simple relation $\sin \mu = 1/M_1$ to obtain the upstream Mach number $M_1$ from the measured Mach angle $\mu$ \citep{Kundu-2012}. Using this relation, we estimate the upstream Mach number to be $M_1 = 5.2 \pm 0.3$ from the Mach angle measured in the experimental end-on image, and $M_1 = 8.2 \pm 0.6$ from the Mach angle in the experimental side-on image. 

 In a hydrodynamic shock, the sound wave, which propagates isotropically at the ion sound speed $C_S$, sets the Mach number \citep{goedbloed_keppens_poedts_2010}. However, in a fast magnetohydrodynamic shock, the fast magnetosonic wave determines the shock dynamics, and this wave has an anisotropic phase velocity --- the wave speed is largest in the direction perpendicular to the magnetic field and smallest in the direction parallel to it. Anisotropy in the fast wave phase velocity leads to anisotropy in the Mach angle, however this anisotropy is small in the high $\beta$ $(V_{FMS} \approx C_S)$ and low $\beta$ $(V_{FMS} \approx V_A)$ regimes \citep{SpreiterStahara1985,Verigin2001}.
 
 We can estimate the Alfvén speed $V_A$ and the Alfvén Mach number $M_A$ upstream of the probe from the electron density and magnetic field measurements. At 300 ns after current start, the line-averaged electron density just upstream of the probe (5 mm from the wires) is $n_e \sim 1.2 \times 10^{18} \SI{}{\per \centi \metre \cubed}$. Assuming an average ionization $\bar{Z} \sim 3.5$ \citep{Suttle_2019}, we find the that upstream mass density is $\rho \sim 1.6 \times 10^{-5} \SI{}{ \gram \per \centi \metre \cubed}$. Combining this with the measured magnetic field $B =  10.6 \pm 2$ T at $t = \SI{300}{\nano \second}$ (and assuming that the magnetic field is unperturbed by the shock), we estimate the upstream Aflvén speed to be $ V_A = 67 \pm \SI{13}{\kilo \metre \per \second}$. Furthermore, we infer the flow velocity $V = 74 \pm \SI{14} {\kilo \metre \per \second}$ from the time delay in the probe signals \citep{Datta2022}, giving us an estimated upstream Alfvén Mach number of $M_A \sim 1.1 \pm 0.3$.  Similarly, for a $T_e \sim T_i \sim 10$ eV aluminum plasma \citep{Suttle_2019}, the sound speed is approximately $C_S \sim \sqrt{\gamma_\text{{eff}}(T_i + Z T_e) / m_i} \sim \SI{12}{\kilo \metre \per \second}$, and the sonic Mach number is, therefore, $M_S \sim 6$. These values are consistent with experimental results from aluminum plasmas in similar exploding wire arrays which show that the fast and Alfvén Mach numbers are expected to be approximately $\sim 2$, while the sonic Mach number is $M_S > 5$ \citep{Burdiak2017}. These values of the sonic and Alfvén Mach numbers indicate that the fast magnetosonic speed $V_{FMS}$ is approximately equal the Alfvén speed $V_A$ (and $\beta$ is small), so the anisotropy in the fast wave speed is small. The expected value of the upstream Mach number determined from the shock geometry ($M_S \sim 5-8$) is in close agreement with the sonic Mach number, which suggests that the bow shock is hydrodynamic, as opposed to being MHD. 
 
 The ideal MHD Rankine-Hugoniot shock jump conditions reduce to those of a hydrodynamic shock when the magnetic field is small, or when the upstream magnetic field is parallel to the shock normal \citep{goedbloed_keppens_poedts_2010}. Neither of these conditions are satisfied by our plasma --- the magnetic field is dynamically significant $\beta \sim 0.1-1$ \citep{Burdiak2017,Suttle_2019}, and the upstream magnetic field is perpendicular to the shock front at the apex of the obstacle.  In resistive MHD, however, finite resistivity breaks the frozen-in condition of ideal MHD, and the magnetic field may diffuse independently of the plasma velocity. The decoupling of the plasma and magnetic field occurs at the resistive diffusion length $l_\eta$ which makes the magnetic Reynolds number of order unity, i.e. $R_M = Ul_\eta/\bar{\eta} \sim 1$.  Diffusion dominates at length scales smaller than the resistive diffusion length scale $l_\eta$, which may explain the hydrodynamic nature of the observed bow shock. Using characteristic values of $n_e \sim 1 \times 10^{18}$ \SI{}{\per \cubic \centi \metre}, $V \sim$ \SI{75}{\kilo \metre \per \second}, and $T_e \sim 10-15 \text{ eV}$, and using Spitzer resistivity, we estimate a resistive diffusion length of $l_\eta \sim 0.4-0.7$ \SI{}{\milli \metre}, which is comparable to the size of the obstacle. Therefore, under these plasma conditions, the magnetic field can decouple from the plasma flow at the shock, which is consistent with the observed shock structure matching the sonic, rather than magnetosonic, Mach number \citep{Russell2022}. 

The bow shock has a larger opening angle in the end-on plane, which is tangential to the magnetic field, than in the side-on plane, which is orthogonal to the field, which indicates that the magnetic field may introduce anisotropy into the shock structure. One possible cause of the observed anisotropy could be magnetic draping. When magnetic field lines frozen into the flow approach an obstacle, they may pile-up ahead of the obstacle, slip past it, or diffuse through the obstacle (including the thin layer of stagnated plasma on the obstacle surface). The rate of pile-up depends on the relative rates of advection and resistive diffusion. If the rates of advective slipping and diffusion are small, then the magnetic field will drape around the obstacle, and the magnetic tension of the bent field lines will provide an additional force opposing the ram pressure of the incoming upstream flow. This will result in a larger opening angle and stand-off distance of the shock \citep{Burdiak2017}. In the end-on plane, the curvature of bent fields lines is expected to increase the shock's opening angle. However, the radius of curvature of the field lines does not lie in the side-on plane, so the bending of field lines does not affect the side-on shock angle. Therefore, we expect magnetic draping to modify the shock geometry only in the end-on plane and not on the side-on plane. We further investigate the effect of flux pile-up and field line draping on the shock structure in the following section. 

\subsection{Comparison with Simulations}

\subsubsection{Shock Structure and Flow Properties}

\begin{table*}\centering
\ra{1.3}
\caption{Summary of measured values and comparison with simulation}
\begin{tabular}{rccc}
\hline
 & & Experiment & Simulation \\
\hline
Opening half-angle (\SI{}{\degree}) & End-On ($B_\parallel$) & $30 \pm 0.5$ & $13 \pm 0.3$ \\
 & Side-On ($B_\perp$) & $7 \pm 0.5$ & $7.6 \pm 0.3$ \\
\hline
Mach angle (\SI{}{\degree}) & End-On ($B_\parallel$) & $11 \pm 0.5$ & $7.4 \pm 0.3$ \\
 & Side-On ($B_\perp$) & $7 \pm 0.5$ & $7.6 \pm 0.3$ \\
 \hline
 Mach No.  & End-On ($B_\parallel$) & $5.2 \pm 0.6$ & $7.8 \pm 0.3$ \\
 (from geometry) (-)& Side-On ($B_\perp$) & $8.2\pm 0.6$ & $7.6 \pm 0.3$ \\
\hline
Sonic Mach No. (-) & & $\sim 6$ & $7 \pm 1$ \\
\hline
Alfvén Mach No. (-) & & $1.1 \pm 0.3$ & $1.2 \pm 0.2$\\
\hline
\end{tabular}
\label{tab:compare}
\end{table*}

Figure \ref{fig:sim_endon} shows that the simulated upstream flow is qualitatively similar to the experimentally observed flow. In the end-on plane, the upstream flow in both the simulated and experimentally observed electron density maps is modulated in the azimuthal direction due to the formation of oblique shocks. In the side-on plane, the upstream flow is relatively more uniform, and the electron density decreases with distance from the wires. Table \ref{tab:compare} provides a summary of the experimentally measured and simulated values relevant to the shock physics. The simulated upstream sonic and Alfvén Mach numbers are in good agreement with the experimental values. Here, we calculate the sonic and Alfvén speeds using $C_S = \sqrt{\gamma_\text{eff} p / \rho}$ and $V_A = B / \sqrt{\mu_0 \rho}$ respectively, where $p$ is the thermal pressure, $\rho$ is the mass density, $\gamma_\text{eff}$ is the effective adiabatic index, and $B$ is the magnetic field strength just upstream of the shock. Here, we use a $\gamma_\text{eff} = 1.13$ for the calculation \citep{Drake2006}. Finally, the opening angle of the shock in the side-on plane also agrees well with the that in the experiment. The Mach number obtained from the shape of the simulated bow shock agrees with the upstream sonic Mach number, showing that the shock is hydrodynamic, just like in the experiment.

The simulated bow shock, however, exhibits a smaller opening angle in the end-on plane than in the experiment. When we account for the direction of the upstream velocity vector, the shock angles of the simulated shock approach similar values in the end-on and side-on planes. This shows that the simulated shock under-predicts the anisotropy observed in the experiment. 

\subsubsection{Effect of magnetic draping and shock anisotropy}

\begin{figure}
\includegraphics[width=1.0\textwidth,page=5]{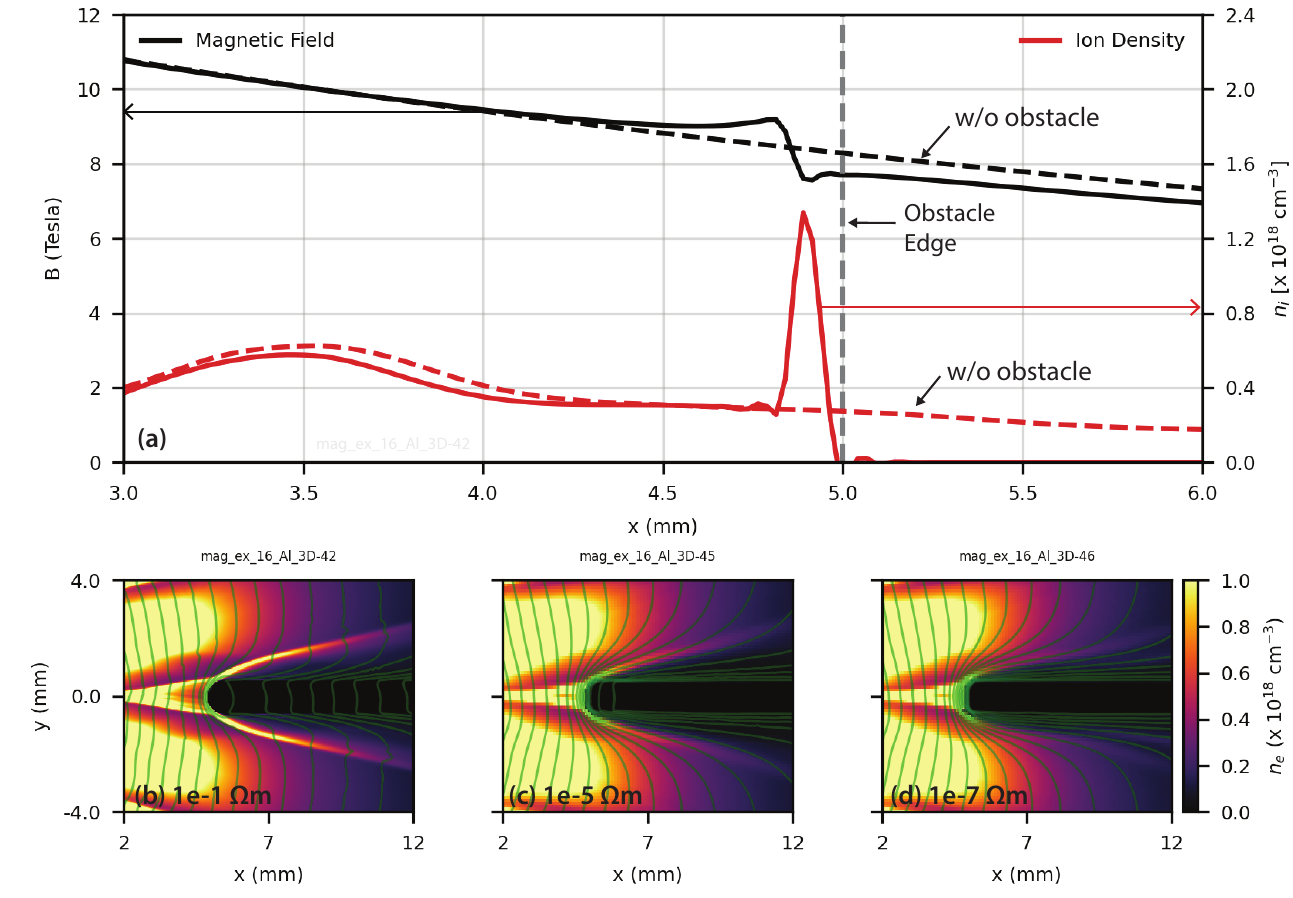}
\centering
\caption{(a) Line-outs of the simulated magnetic field and electron density along the obstacle-axis. (b - d) End-on electron density with overlaid magnetic field lines for obstacle resistivities $10^{-1}\SI{}{\Omega \metre}$, $10^{-5}\SI{}{\Omega \metre}$, and $10^{-7}\SI{}{\Omega \metre}$ respectively.
}
\label{fig:sim_B_field}
\end{figure}

A line-out of the magnetic field along the obstacle axis shows that the magnetic field increases by $\sim$ 1.5 T at the shock front (Figure \ref{fig:sim_B_field}a). Note that the jump in the magnetic field is much smaller than that in ion density ($\sim 4\times)$, suggesting that the rise in field strength is due to flux pile-up at the obstacle, rather than due to shock compression, which would result in comparable jumps in the magnetic field and ion density. The pile-up occurs because the obstacle and the layer of stagnant plasma ahead of it have some finite conductivity, which limits the rate of resistive diffusion. Figure \ref{fig:sim_B_field}b shows magnetic field lines overlaid on the simulated electron density. The field lines drape around the probe, but are able to diffuse through the resistive obstacle, which limits the amount of flux pile-up. Note that in the experiment, the leading edge of the probe consists of loops of copper wire coated in insulating enamel.

To investigate the importance of magnetic draping, we repeat the simulations with obstacles of increasing conductivity, as shown in Figures \ref{fig:sim_B_field}c ($10^{-5}\SI{}{\Omega \metre}$) \& \ref{fig:sim_B_field}d ($10^{-7}\SI{}{\Omega \metre}$). Increasing the obstacle conductivity leads to increased flux pile-up and magnetic draping; however, it does not generate a wider bow shock. This suggests that some mechanism, other that magnetic draping, may contribute to the experimentally-observed anisotropy in the shock shape. 

One possible mechanism that can create anisotropic hydrodynamic shock structures is pressure anisotropy, which will generate ion acoustic waves that travel at different speeds along and across magnetic field lines \citep{Hau1993,Chou2004}. Allowing for an anisotropic pressure tensor ${\bf P} = p_\perp {\bf I} + (p_\parallel - p_\perp) {\bf bb}$, we estimate that in order to obtain the observed anisotropy in the bow shock shape, we require $p_\parallel \sim 2.5 p_\perp$. Significant pressure anisotropy, however, is unlikely in our plasma, given that it is highly collisional, and has a pressure-anisotropy relaxation time  $\sim \SI{0.1}{\nano \second}$ \citep{Huba2013}, which is small compared to the hydrodynamic time-scale of the experiment. 

 Note that the ion skin depth $d_i \sim \SI{0.5}{\milli \metre}$ is also comparable to the resistive diffusion length in this plasma. When ions and electrons decouple at the ion skin depth, the magnetic field is no longer frozen into the ion fluid, and two-fluid effects, in particular the Hall term in Ohm's law, may become important \citep{Eastwood2007,Shaikhislamov2013}. Extended-MHD simulations have previously demonstrated widening of MHD bow shocks around obstacles with size comparable to the ion inertial length, and this modification of the shock shape is associated with a suppression of the current in the post-shock low density region by the the Hall term ${\bf j \times B}/en$ \citep{Zhao2015}. However, it remains unclear if Hall effects can account for the observed shock anisotropy in the experimental bow shock, and future work will aim to investigate this with extended-MHD simulations.

In the analysis above, we have assumed a constant $\gamma_\text{eff}$ of the plasma. However, since the magnetic field remains frozen-into the electrons, which have different degrees of freedom $n$ along ($n = 1$) and across ($n = 2$) the magnetic field, the value of $\gamma$ can also exhibit anisotropy relative to the field direction. Using the relation $\gamma = (n+2)/n$ for adiabatic compression \citep{Chen1974}, we estimate the parallel and perpendicular Mach numbers to be $M_\parallel (\gamma_\parallel = 3) \sim 4$ and $M_\perp (\gamma_\perp = 2) \sim 5$ respectively. The Mach number is larger in the $B_\perp$ plane, and this would lead to a smaller Mach angle in the side-on $B_\perp$ plane ($\mu_\perp \sim \SI{12}{\degree}$)  compared to that in the end-on $B_\parallel$ plane ($\mu_\parallel \sim \SI{15}{\degree}$). This qualitatively reproduces the the anisotropy in the experimental result. A quantitative comparison would require a measurement of the ion and electron temperatures, which can be accomplished in future experiments via optical Thompson scattering (OTS) \citep{Suttle2021}.
 
 Finally, in calculating the Mach angle from the experimental shock opening angle, we assumed a radial velocity field. In reality, the flow may also exhibit an azimuthal component of velocity due to the thermal expansion of the plasma ablating from the wire cores \citep{Swadling2013}. This may cause the calculated value of the Mach angle in the end-on plane to be an underestimate of the true value. OTS along two lines of sight can be employed to better estimate the angle of the upstream velocity vector \citep{Suttle2021}.

\subsubsection{Magnetic Field Measurements}

Figure \ref{fig:bdot_measuremnts}b shows that the simulated magnetic field at the probe location 300 ns after current start is 8 T, which is $\sim 8-30 \%$ lower than the experimentally measured field of $10.6 \pm 2$ T at the same time. 
The peak simulated field at the probe occurs at $t \sim \SI{300}{\nano \second}$, and is weaker than the experimentally observed peak field ($11.5 \pm 2.5$ T), which occurs later at $t \sim \SI{340}{\nano \second}$. However, when the simulation is repeated with a peak current of $I_{pk} = 1.4$ MA, which is the upper bound on the current delivered to the load, the simulated magnetic field at $t = \SI{300}{\nano \second}$ is 11 T, which is in better agreement with the experimentally measured field. We reiterate, however, that the actual current delivered to the load is likely to be smaller than 1.4 MA, given that the wire array does not explode in the experiment. 

The experiments and simulations independently show that under the given plasma conditions, the shock around the b-dot probe is hydrodynamic. The simulations additionally show that the effect of magnetic flux pile-up on the measured field is expected to be small, which suggests that the b-dot probe does not significantly perturb the magnetic field. However, the measured magnetic field is larger than the simulated field, and further investigation, with an array of probes fielded over repeated shots to better determine the uncertainty in the measured magnetic field, is required to resolve this discrepancy. We also note that since the probes were calibrated using a slower $\sim \SI{1}{\micro \second}$ current pulse compared to MAGPIE's $\SI{250}{\nano \second}$ rise time current response, there may be additional unaccounted systematic error in the probe calibration. To ascertain the systematic error in the b-dot probe measurement, simultaneous magnetic field measurements alongside other magnetic diagnostics, such as Zeeman splitting of spectral lines \citep{Rochau2010}, or Faraday rotation polarimetry \citep{Swadling2014}, can be performed in future work. 

\section{Conclusions and Future Work}

We have presented experimental results and numerical studies of 3D bow shocks generated in a collisional magnetized plasma from the interaction of a pulsed-power-driven supersonic ($M_S \sim 7$) super-fast magnetosonic ($M_{FMS} > 1$) flow with a small inductive probe. Line-integrated electron density obtained from imaging interferometry 300 ns after current start shows a well-defined detached bow shock ahead of the probe. The bow shock exhibits a fully 3D anisotropic structure, with a larger opening angle in the end-on plane (parallel to the magnetic field) than in the side-on plane (perpendicular to the magnetic field). Since the resistive diffusion length in the plasma is comparable to the size of the probe, the magnetic field decouples from the ion fluid, and we expect the bow shock to be hydrodynamic, rather than MHD. From the shock geometry, we estimate the upstream Mach number ($ 5 < M_1 < 8$) of the flow, and find the calculated Mach number to be in good agreement with the sonic Mach number of the plasma, showing that the shock is hydrodynamic. 

We compare our experimental results with fully 3D resistive MHD simulations of the experimental setup using the code GORGON, which confirms the hydrodynamic nature of the shock. The simulation successfully reproduces several features of the experiment, including upstream density modulation, the magnitudes of the sonic and Alfvénic Mach numbers, as well as the shock opening angle in the plane perpendicular to the magnetic field. The simulation, however, under-predicts the anisotropy observed in the shape of the experimental shock. We explore multiple possible mechanisms that can introduce anisotropy in the bow shock shape. The simulations show that the shock shape in this diffusion-dominated regime remains largely independent of magnetic flux pile-up and magnetic draping. Next, we also find that the pressure anisotropy required to generate the observed anisotropy in the shock shape is large, and unlikely to occur in our plasma since the pressure anisotropy relaxation time is small compared to the hydrodynamic time scale of the plasma. Anisotropic $\gamma$ due to different degrees of freedom of the electron fluid along and across the magnetic field qualitatively describes the experimentally-observed shock anisotropy, but a quantitative comparison requires a measurement of the temperature. Finally, since the resistive diffusion length scale is comparable to the ion inertial length, the Hall term may become important. Future work will focus on the importance of the Hall term, via extended-MHD simulations of the bow shock. 


Future experimental work will also attempt to quantify the change in the direction and magnitude of the velocity vector, as well as in density and pressure across the shock front. This can be implemented using the ion feature of optical Thomson scattering, which can provide independent measurements of the flow velocity, ion and electron temperatures, and the ion sound speed.

\section{Acknowledgements}

This work was funded in part by NSF and NNSA under grant no. PHY2108050, and supported by the U.S. Department of Energy (DOE) under Award Nos. DE-SC0020434, DE-NA0003764, DE-F03-02NA00057, DE-SC-0001063, and DE-NA0003868, and the Engineering and Physical Sciences Research Council (EPSRC) under Grant No. EP/N013379/1. The simulations presented in this paper were performed on the MIT-PSFC partition of the Engaging cluster at the MGHPCC facility (www.mghpcc.org) which was funded by DOE grant no. DE-FG02-91-ER54109.

\bibliographystyle{jpp}
\bibliography{jpp_instructions.bib}

\end{document}